\documentclass[]{rsos}%%%%where rsos is the template name
\usepackage{graphicx}
\usepackage{amsmath}
\usepackage{xcolor}
\usepackage{tcolorbox}
\usepackage{enumitem}
\usepackage{authblk}

\begin{document}

%%%% Article title to be placed here
\title{Structure-based Hamiltonian model for IsiA uncovers a highly robust pigment protein complex }

\author{%%%% Author details
Hanan Schoffman$^{1,a}$, William M. Brown$^{2,a}$, Yossi Paltiel$^{3}$, Nir Keren$^{1}$ and Erik M. Gauger$^{2}$}

%%%%%%%%% Insert author address here
\address{$^{1}$Department of Plant and Environmental Sciences, The Alexander Silberman Institute of Life Sciences, The Hebrew University of Jerusalem, Jerusalem 91904, Israel\\
$^{2}$SUPA, Institute of Photonics and Quantum Sciences, Heriot-Watt University, Edinburgh EH14 4AS, UK\\
$^{3}$Applied Physics Department, The Hebrew University of Jerusalem, Jerusalem 91904, Israel\\
$^{a}$Joint first authorship.}

%%%% Subject entries to be placed here %%%%
\subject{Biophysics, Computational biology, Bioenergetics}

%%%% Keyword entries to be placed here %%%%

%%%% Insert corresponding author and its email address}
\corres{Nir Keren and Erik M. Gauger\\
\email{nir.ke@mail.huji.ac.il}\\
\email{e.gauger@hw.ac.uk}}

%%%% Abstract text to be placed here %%%%%%%%%%%%
\begin{abstract}
The iron stress-induced protein A (IsiA) is a source of interest and debate in biological research. The IsiA super-complex, binding over 200 chlorophylls, assembles in multimeric rings around photosystem I (PSI). Recently, the IsiA-PSI structure from \emph{Synechocystis} sp.~PCC 6803 was resolved to 3.48 {\AA}. Based on this structure, we created a model simulating a single excitation event in an IsiA monomer. This model enabled us to calculate the fluorescence and the localisation of the excitation in the IsiA structure. To further examine this system, noise was introduced to the model in two forms -- thermal and positional. Introducing noise highlights the functional differences in the system between cryogenic temperatures and biologically relevant temperatures. Our results show that the energetics of the IsiA pigment-protein complex are very robust at room temperature. Nevertheless, shifts in the position of specific chlorophylls lead to large changes in their optical and fluorescence properties. Based on these results we discuss the implication of highly robust structures, with potential for serving different roles in a context dependent manner, on our understanding of the function and evolution of photosynthetic processes.
\end{abstract}
%%%%%%%%%%%%%%%%%%%%%%%%%%%

%%%%%%%%%%%%%%% End of first page %%%%%%%%%%%%%%%%%%%%%

\maketitle

\section{Introduction}

Bioenergetic processes may be perceived as being fine-tuned for maximum efficiency through selective pressures towards the highest possible yields. However, it should be taken into account that the concept of fitness refers to an organism rather than a single trait. Therefore, evolution might select for plasticity over yield in an ever-changing environment that creates the need for fast acclimation responses. Photosynthesis is arguably the most important bioenergetic process on Earth. The structures of many of the pigment-protein complexes involved in this interactive and tightly regulated process are well conserved throughout the evolutionary tree~\cite{Mazor2015,Hohmann-Marriott2011,Pandit2018}.
The study of the photosynthetic machinery is progressing through biochemical, molecular biology and biophysical means, with major breakthroughs being owed to the elucidation of their structures~\cite{Rutherford2001,Nelson2006}. One aspect that is still hard to resolve based on structural data and experimental methods, is whether and how biological systems, which have evolved to operate at temperatures of around 300~K, use vibrational modes and how these influence function~\cite{Lambert2013, Chin2013}. Another facet of relevance to experimental methodology is that a small change in a protein's molecular structure -- phosphorylation or protonation, for example -- is enough to exert large changes on its three-dimensional structure and greatly affects its function. Both structural changes and thermal noise are undoubtedly present in photosynthesis and can potentially either be taken advantage of, or be rendered unimportant by structures which intrinsically mitigate their effects~\cite{Keren2018,Chin2012,Arp2020, Cao2020}.   

In the present manuscript, we probe the hypothesis by which biological structures have evolved to operate under environmental conditions and take  advantage of the ever present noise. We explore this by using a computational approach focusing on the cyanobacterial iron stress-induced protein A (IsiA). The IsiA pigment protein complex is expressed by many fresh water cyanobacteria under iron limitation~\cite{Laudenbach1988,Burnap1993}. Its chlorophyll-binding monomers create a ring encircling photosystem I (PSI). A recently resolved structure of the IsiA-PSI complexes showed 18 identical subunits binding over 200 chlorophyll molecules~\cite{Toporik2019}. Its role in the iron limited cyanobacterial cell has been debated extensively over the past thirty years~\cite{Berera2009,Bibby2001,Chen2018,Chen2017,Havaux2005,Ihalainen2005,Ryan-Keogh2012,Schoffman2019,Singh2007,Sun2015,Yeremenko2004}.

A key issue in the study of IsiA's function is the role this complex has in energy transfer to PSI. The two main hypotheses regarding IsiA's function in energy transport are: (a) An auxiliary antenna for PSI which compensates for the loss of high iron demanding PSI units by increasing cross section. (b) Excitation energy quenching, preventing excess energy from reaching PSI. In a nutrient limited organism, excess redox power can result in the generation of damaging reactive oxygen species. Both proposed functions accumulated evidence in research, with several studies providing support for a dual functionality of the IsiA complex in a cell-physiology context dependent manner~\cite{Yeremenko2004}. The debate over the biological function of IsiA makes it interesting to probe, computationally, the extent of the perturbation that is needed to switch the energy transfer properties of this protein. 

\begin{figure}
\centering
\includegraphics[width=0.9\textwidth]{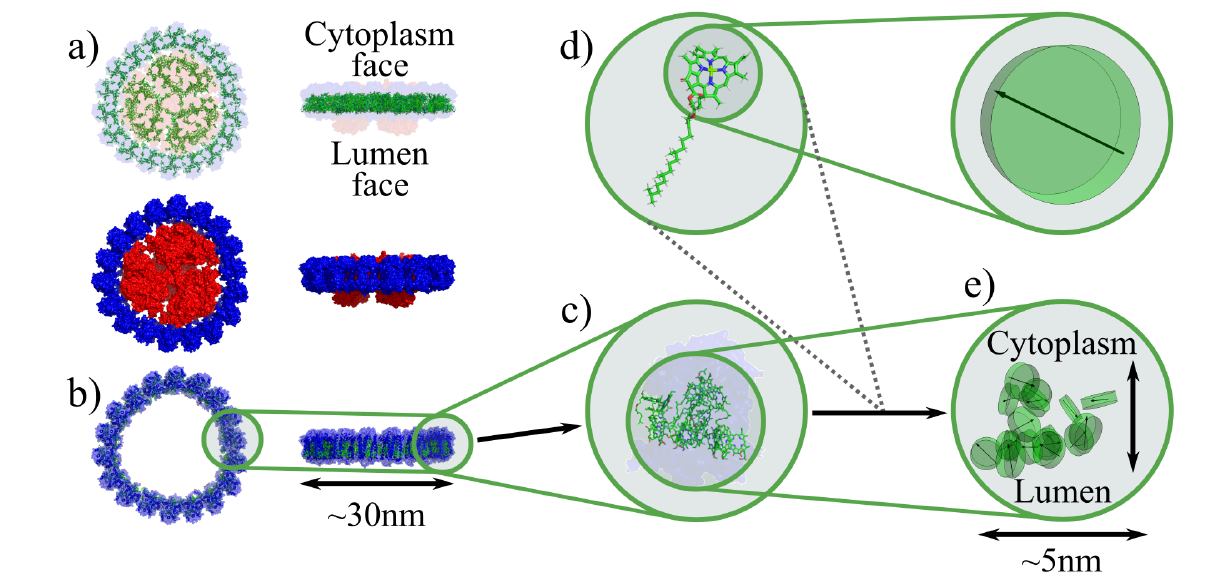}
\caption{Schematic of how we produce a tractable system to model from the complicated experimental data of the structure of IsiA. a) Full IsiA-PSI rings including all pigments. The top view shows the chlorophylls highlighted in green, whereas the bottom view separates the PSI from the IsiA components, shown, respectively, in red and blue. b) The IsiA structure without the central PSI. c) We consider the chlorophyll A molecules of a single IsiA monomer subunit, ignoring other pigments. d) We ignore the spectroscopically unimportant chlorophyll A tails and model the chromophores as cylindrical disks to arrive at our model structure displayed in e).}
\label{fig:model_progression}
\end{figure}

The newly resolved structure of IsiA-PSI (PDB entry 6NWA, Ref.~\cite{Toporik2019}), shown in Fig.~\ref{fig:model_progression}a, creates an opportunity for a new approach to clarify the issue of IsiA's function. As a position was ascribed for each chlorophyll and since the physical properties of chlorophyll are well known, a model of the system can be created and manipulated to test energetic scenarios not attainable via experimental means. In this work, we create such a computational Hamiltonian model, comprising a network of coupled chlorophylls to study the IsiA's excitonic properties following an excitation event. 

The close proximity of the chlorophylls in the system begs the question whether this system might display non-trivial quantum mechanical effects of the type attributed to other photosynthetic structures in the context of quantum biology~~\cite{Lambert2013,Chin2013,Romero2014}. However, motivated by finding that the exciton lifetimes of IsiA~\cite{Berera2009} exceed plausible timescales for electronic coherence~\cite{Lambert2013}, this study does not consider `dynamic' quantum coherent effects (with respect to the energy basis). As will become apparent shortly, we do consider the properties of spatially delocalised states (thus involving `static' coherence in the site basis).

\section{Model}

\begin{tcolorbox}

Key physics concepts for biologists:

\begin{description}[align=left]
\item [Eigenstate:] Our model builds upon a `site basis' picture of individual chlorophylls, assigning on-site (transition) energies for each chlorophyll molecule and off-diagonal coupling terms which allow population to transfer between molecules. When we diagonalise the resulting system Hamiltonian, we move to a basis known as energy-, excitonic- or simply eigenbasis. 

Eigenstates form a complete set, and each is a linear combination of site basis states. Each eigenstate has an associated energy, or eigenvalue, which is determined by the site energies and coupling terms from the site basis. An isolated system in a pure eigenstate will remain indefinitely in that state, otherwise it will transfer population between eigenstates. Optical excitations on the network are delocalised and best represented in the eigenbasis.
\item [Brightness/darkness:] We use the terms `brightness' and `darkness' to, respectively, refer to single-photon super- and sub-radiance: when one has a collection of identical optical dipoles in a specific geometric arrangement, one can measure the joint optical emission rate. If this rate is greater (less) than the rate for one of the individual dipoles in isolation, then one can say that the arrangement of dipoles is bright (dark). In other words: a collection of emitters which holds onto an excitation for longer than a single molecule is dark, but if it emits it faster then it is considered bright. 
\item [Leaking value:] We introduce the `leaking value' ($L$) as a normalised scalar measure of how bright or dark an arrangement of molecules is. Specifically, it indicates the oscillator strength for the expected decay process of a single excitation in the complex. For a system with $N$ identical dipoles the value falls in the range $0\leq L\leq N$, where $L>1$ represents a system that is exhibiting brightness, meaning it will emit an excitation faster than an isolated emitter. Conversely, $L<1$ represents a system that is exhibiting darkness, meaning it will emit an excitation slower than an isolated emitter. A bright system with $L=N$ would relax $N$ times faster than an isolated single dipole, whereas a dark system where $L=0.1$ would decay ten times slower than an isolated molecule.
\end{description}

\end{tcolorbox}

\subsection{Dipole geometry}

\begin{figure}
\centering
\includegraphics[width=\textwidth]{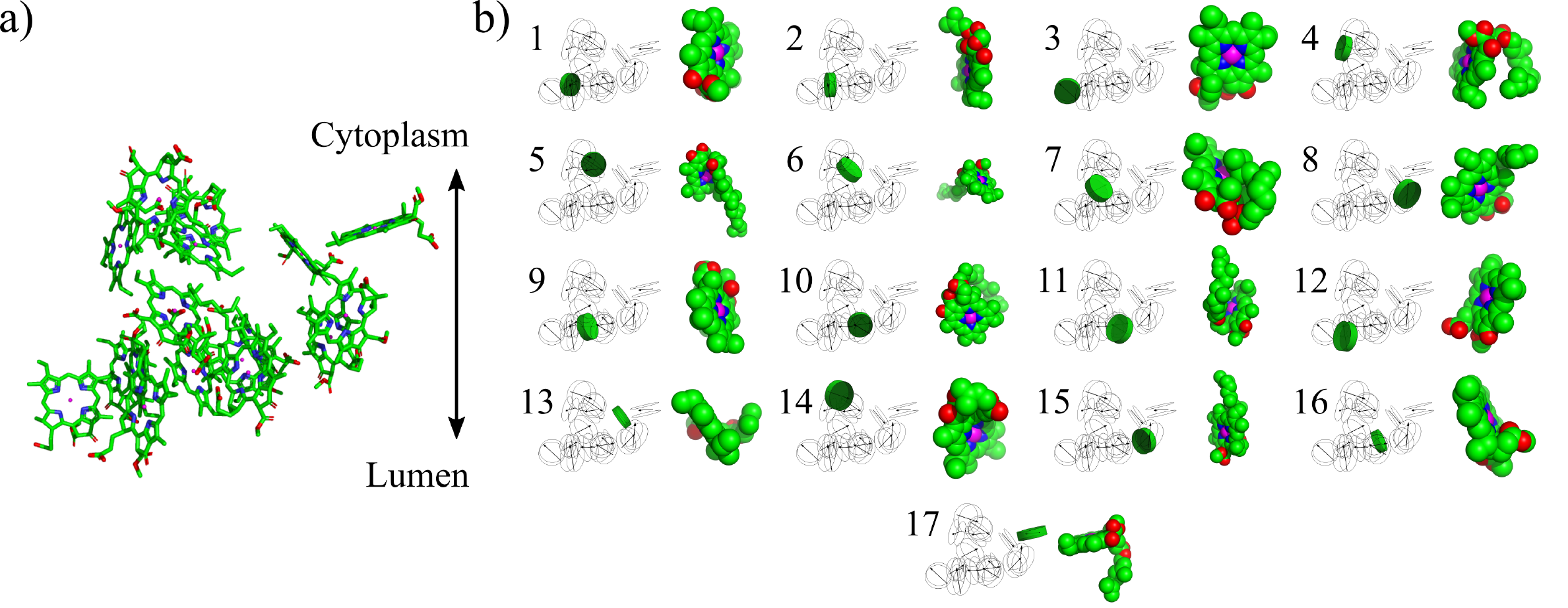}
\caption{We here show the positions of the different chlorophyll A molecules in the IsiA monomer. a) gives a `master view' of our structure including its orientation in the organism. b) Each of the 17 subpanels highlights the position of a single chlorophyll A molecule (represented as a disk) in the monomer on the left. A space-filling molecular ball representation of the molecules is shown on the right of each panel, in the orientation as seen from the PSI trimer centre of an IsiA ring. The reference atoms for the disk approximation of the structure (see text) are magnesium (pink), and nitrogen (blue). The assignment of numerical labels of each site within the IsiA monomer follows that of Ref.~\cite{Toporik2019}.}
\label{fig:ChlorophyllA_wiki}
\end{figure}

Figure~\ref{fig:model_progression} gives a step by step depiction of how we extract a tractable geometrical model from the full ring system found by Toporik et al~\cite{Toporik2019}: First, we drop the central reaction centre and only concern ourselves with the ring of IsiA monomers surrounding it. Second, we disregard all of the pigments and proteins except for the chlorophyll A molecules of a single IsiA monomer. Carotenoids are also neglected here, in line with many other photosynthetic antenna modelling efforts due to their complex excitonic structure~\cite{Hashimoto2016, Larsen2003}.
Finally, we approximate individual molecules as disks. These choices are explained in more detail in the following.

Three identical groups of six such monomers repeat around PSI in Fig.~\ref{fig:model_progression} to produce a full ring. While we could model the full ring system, interactions between monomers are relatively weak and the closer separations within each monomer contribute the dominant excitonic coupling terms. Therefore, we here focus on a single IsiA monomer comprised of seventeen chlorophylls, and expect this will capture the most significant properties and behaviour of the system. This reduction has the added advantage of keeping the dimension of our model smaller and allowing for an easier interpretation and visualisation of results.

The molecular positions and orientations of the dipoles derive from the cryo-EM results measured by Toporik et al~\cite{Toporik2019}. From hereon, we refer to these cryo-EM positions as our `master' configuration. The individual chlorophylls are numerically labelled and their positions within the IsiA monomer are shown in Fig.~\ref{fig:ChlorophyllA_wiki}; our numbering convention matches that of Ref.~\cite{Toporik2019}. Wishing to introduce shifts and disorder to the molecular positions, we need to ensure chlorophylls do not begin overlapping when they become too close. The simplest approach would be assuming a spherical exclusion volume for each molecule, but considering the actual structure, a smaller `disk' approximation will be more accurate. For simplicity we ignore the `tail' appendix, as it is not rigid and instead acts as a flexible hydrophobic anchor and does not effect the spectral properties of the chlorophyll, see Fig.~\ref{fig:ChlorophyllA_wiki}.

\begin{figure}
\centering
	\includegraphics[width=0.6\textwidth]{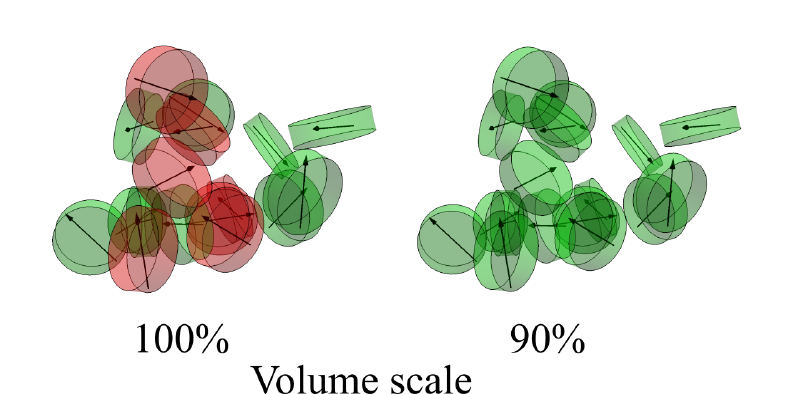}
\caption{An IsiA monomer with two different scales for approximated volume profiles. Chlorophyll A molecules which intersect with the volume of another are highlighted red. The dipole directions of each chlorophyll are denoted by black arrows.}
\label{fig:monomeroverlap}
\end{figure}

The following paragraph is concerned with technical subtleties regarding our choice of disks and avoiding volume overlap. It has been included to fully describe our approach, but is not necessary for understanding and appreciating our results discussion and conclusion.
To fix the dimension of the chlorophyll disks, we set the thickness to the van-der-Waals radius of the central magnesium atom (0.173~nm). Were we to follow Ref.~\cite{Toporik2019} and use a chlorophyll diameter of 1.23~nm, we would see overlap even in the master configuration. The left panel of Fig.~\ref{fig:monomeroverlap} highlights the overlapping chlorophylls (6, 7, 10, 11, 12, 14, and 16) in red. In order to avoid overlap in our master positions we could use a more complicated exclusion volume, for example a tapered disk with reduced thickness further away from the centre (to account for other atoms having smaller van-der-Waals radii than magnesium). However, this would significantly increase the computational complexity of our overlap checks. Instead we choose to decrease the disk dimensions by 10\%, allowing us to maintain the simpler shape profile. On the right-hand side of Fig.~\ref{fig:monomeroverlap} we see that this resolves the issue of overlaps in the master configuration. 

\subsection{System Hamiltonian and effective model }

We wish to model the chlorophyll A molecules in IsiA as a collection of optical dipoles, each represented as a two-level system (2LS), following numerous theoretical studies of photosynthetic complexes~\cite{Grondelle1994, Ishizaki2009}. Each 2LS has an energy splitting, $\omega_\alpha$, as well as a dipole moment, $\textbf{d}_\alpha$. We note that the decision to model as a repeated single 2LS with a unique site energy assumes uniformity in the fluorescence and absorption peaks; this is unlikely to be strictly true, however, we do not expect our conclusions to be significantly affected by this simplification. The molecule's radiative lifetime, $\tau_{\alpha} = 15$~ns, \cite{Mar1972, Rabinowitch1957}
is used alongside $\omega_\alpha$ = 1.85~eV to determine the magnitude of the dipole moment, $|\textbf{d}_\alpha| = \sqrt{3\pi \epsilon_0 \tau_{\alpha}^{-1} c^3 / \omega_\alpha^3}$~\cite{Agarwal1974,Shatokhin2018}. Since the distances between chlorophylls are on the order of single nanometres~\cite{Toporik2019}, they are small relative to the optical wavelengths, meaning the whole system interacts collectively with the surrounding optical field~\cite{Breuer2002}. Their proximity also induces resonant dipole-dipole F\"{o}rster-type couplings between different sites~\cite{Varada1992,Curutchet2017}. The coupling, $J_{\alpha,\beta}$, between two chlorophylls depends on both the dipole moments, and separation vector, $\textbf{r}_{\alpha,\beta}$, taking the form
\begin{equation}
J_{\alpha,\beta}(\textbf{r}_{\alpha,\beta})=\frac{1}{4\pi\epsilon_0 |\textbf{r}_{\alpha,\beta}|^3}\bigg(\textbf{d}_\alpha\cdot\textbf{d}_\beta-\frac{3(\textbf{r}_{\alpha,\beta}\cdot\textbf{d}_\alpha ) ( \textbf{r}_{\alpha,\beta}\cdot\textbf{d}_\beta)}{|\textbf{r}_{\alpha,\beta}|^2}\bigg) ~.
\label{eq:forster}
\end{equation}
This expression for the dipole-dipole couplings is effective for point dipoles, so for it to be accurate for our molecular dipoles there needs to be adequate separation between the dipoles for the approximation to hold, otherwise a treatment based on fractional charges becomes more appropriate~\cite{Curutchet2017}, as is well-established for bacterial reaction centres~\cite{Warshel1987}, and the antenna rings of purple bacteria~\cite{Baghbanzadeh2016a}. However, as most pairwise separations of the IsiA have adequate separation between chlorophylls of well over a nanometre, for simplicity we here employ Eq.~\eqref{eq:forster} throughout. \footnote{To justify this approach, we have checked explicitly that allowing for reductions on the order of a factor of two for the most closely spaced chromophores does not affect our qualitative conclusions. Particularly the  room-temperature results are fully robust to these corrections including at a quantitative level.}

We proceed with the generic collective light matter interaction Hamiltonian that is appropriate for our model~\cite{Breuer2002},
\begin{equation}
\hat{H}_{I,\text{opt}}=\sum_{i=1}^N \mathbf{d}_i\hat{\sigma}_i^x \otimes \sum_k f_k \left(\hat{a}_k+\hat{a}_k^\dagger\right) ~,
\label{eqn:INTopt}
\end{equation}
where $\hat{\sigma}_i^x$ is the standard Pauli operator for site $i$, and $f_k$ and $\hat{a}_k^{(\dagger)}$ are, respectively, the coupling strength, and annihilation (creation) operator for the optical mode $k$. As we lack detailed vibrational information, we do not include a microscopically-derived model for the phonons, but we capture the well-known general behaviour of spin-boson type interactions: i.e.~we include vibrationally-assisted transitions between system eigenstates within the same excitation manifold~\cite{Creatore2013,Higgins2017,Brown2019}. For reference, the relevant multi-site spin-boson interaction Hamiltonian is
\begin{equation}
\hat{H}_{I,\text{vib}}=\sum_{i=1}^N \hat{\sigma}_i^z \otimes \sum_q g_{i,q} \left(\hat{b}_{i,q}+\hat{b}_{i,q}^\dagger\right) ~,
\label{eqn:INTvib}
\end{equation}
where $g_{i,q}$ and $\hat{b}_{i,q}^{(\dagger)}$ are, respectively, the coupling strength, and annihilation (creation) operator for the vibrational mode $q$ for the bath linked to site $i$. As the nanometre scale systems we consider are large relative to characteristic phonon length scales, local phonon environments are the most appropriate choice\footnote{Note, however, that instead considering a shared phonon bath with appropriate position-dependent phase factors~\cite{McCutcheon2009, Gauger2008} would not produce a qualitative difference~\cite{Lim2014}.}. We write our environment Hamiltonian terms as baths of oscillators for our shared photon, and local phonon environments, respectively, as $\hat{H}_{B,\text{opt}}$ and $\hat{H}_{B,\text{vib}}$. Our total Hamiltonian now becomes
\begin{align}
\hat{H}_\text{total} &= \omega_\alpha \sum_{i=1}^N \hat{\sigma}_i^z + \sum_{i \neq j}^{N} J_{i,j} \left(\hat{\sigma}_i^+ \hat{\sigma}_j^- + \hat{\sigma}_i^- \hat{\sigma}_j^+  \right) \\ \nonumber
& + \hat{H}_{I,\text{opt}} + \hat{H}_{I,\text{vib}} + \hat{H}_{B,\text{opt}} + \hat{H}_{B,\text{vib}}~, 
\end{align}
where the first line above is the system Hamiltonian $\hat{H}_\text{sys}$.

\begin{figure}
\centering
	\includegraphics[width=1\textwidth]{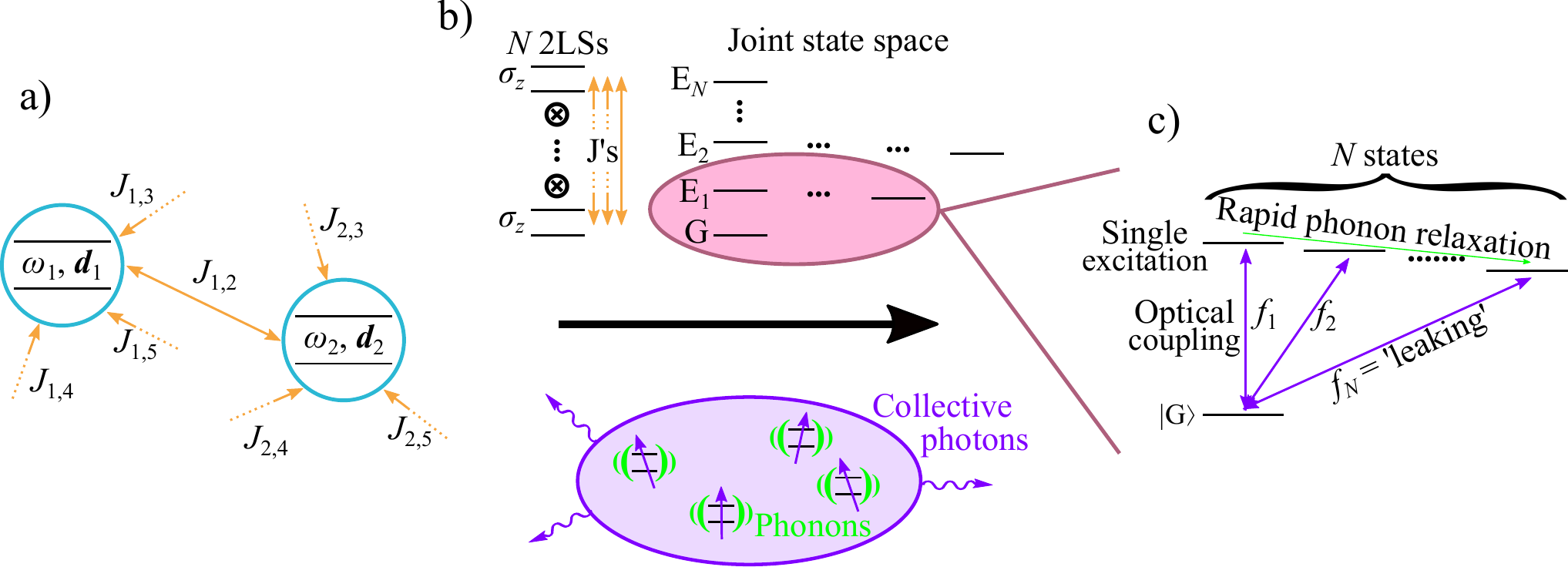}
\caption{Schematic of the reduced system Hamiltonian we are considering. a) Our collection of $N$ dipoles is represented as 2LSs linked by distance and relative orientation dependent coupling terms. b) The site energies and couplings are used to find the eigenbasis, from which we take the eigenstates for the ground state as well as single excitation states. Each 2LS interacts with a collective optical field dependent on its dipole moment. Individual vibrational baths effect each 2LS identically. c) The optical transition strengths and effect of phonon relaxation are then calculated for our chosen eigenstates, allowing us to calculate a leaking value. }
\label{fig:H_schem}
\end{figure}

The dynamics for a small number of chlorophylls can be resolved by treating the interaction terms of this Hamiltonian to second order based on a Born-Markov approximations in the Redfield formalism~\cite{Breuer2002,Brown2019}. However, modelling the full excitation subspace for more than ten chlorophylls with the outlined approach is computationally challenging due to the exponential growth of the Hilbert space of the model, but by limiting ourselves to the overall ground state and single excitation subspace we reduce our Hilbert space for $N$ 2LS molecules from $2^N$ states to $1+N$ states (see Fig.~\ref{fig:H_schem}). 
While multi-excitation effects could boost the performance of man-made absorbing systems~\cite{Higgins2017, Hu2018, Brown2019}, for the low intensity of light that is incident at the surface of the Earth, we do not expect biological light-harvesters to have more than one excitation at any time\footnote{If we assume solar irradiance of 1~kW/m$^2$ with monochromatic light of 500~nm, the interval between incident photons on a cross-section with radius 15~nm would be 0.6~$\mu$s, considerably longer than excitonic lifetimes.}. This simplification also means that we do not need to consider effects such as exciton-exciton annihilation~\cite{Amerongen2000, May2014}.

In Fig.~\ref{fig:H_schem} we present a schematic of our effective model. For an $N$ dipole system we calculate the energies of the $N$ eigenstates of the single excitation manifold by finding the diagonal basis for our system Hamiltonian. We then calculate the dipole contribution for each of the transitions linking single excitation eigenstates to the ground state following the procedure described in Ref.~\cite{Brown2019} and its Supporting Information\footnote{Since all of the IsiA chromophores are identical and the couplings $J_{\alpha,\beta}$ are modest, we assume a flat optical spectral density and neglect any transition frequency dependency on the optical rate~\cite{Breuer2002}.}. Fast vibrational relaxation means that any population in the single excitation manifold will be rapidly redistributed. A common assumption in chemistry is Kasha's rule \cite{Kasha1950}, where all excited state population is expected to occupy the lowest lying state of the single excitation manifold. Relaxation within the single excitation manifold can be modelled by considering phonon-assisted transitions explicitly~\cite{Brown2019,Rouse2019}, finding near-complete transfer into the lowest energy eigenstate in each excitation manifold under suitable conditions.

\subsection{Concept of the leaking value}

We proceed to define the oscillator strength linking the lowest excited state to the ground state as the `lowest-state leaking value', $L$. A leaking value greater (less) than one then represents a system that is exhibiting brightness (darkness), meaning it will emit an excitation faster (slower) than an isolated emitter. As will become apparent in our analysis, we shall be primarily interested in darkness, which implies collective quantum effects are used to hold on to excitations for longer than would be classically expected. (By contrast, brightness could support features such as supertransfer~\cite{Baghbanzadeh2016} and fast radiative quenching, but these are not observed in the current structure.)

Assuming all of the dipoles are identical, as is the case here, with a lifetime $\tau_\alpha$ we can normalise such that a $L>1$ is bright, and $L<1$ is dark. One can then get an indication of the radiative lifetime of an excitation in such a system by considering $\tau_\alpha/L$. By contrast, for a collection of dipoles of differing brightness the choice for normalisation becomes less clear. Nonetheless, as long as there is consistency in defining a benchmark lifetime\footnote{For example, one could benchmark against the darkest contributing chromophore, which will have the longest lifetime, or alternatively use the most common dipole or a weighted average based on the count of each type of molecule present.}, one can still use the concept of a leaking value to compare different geometries of the same collection of chromophores. Note that we here only focus on radiative processes and neglect non-radiative decay processes, which provide other competing recombination channels~\cite{Amerongen2000}.

For energy splittings between eigenstates from the induced F\"{o}rster couplings that are relatively small the assumption that all population will relax to the lowest energy state in the single excitation manifold becomes invalid. If, for example, the lowest energy eigenstate has a very small leaking value, but a state with an eigenenergy just a few meV higher has a strong contribution, then one would still expect the higher energy state to be populated at room temperature ($k_B\times 300$~K $\approx25$~meV). To account for this we introduce the `Boltzmann-corrected leaking value' $L_{\rm boltz}$, where we look at the dipole contributions of every transition linking a single energy excitation eigenstate to the ground state, and sum them with a Boltzmann weighting,
\begin{equation}
w_i=\frac{e^{-E_i\beta}}{\sum_k e^{-E_k\beta}}~,
\end{equation}
where $i$ is the single energy eigenstate, $E_i$ is the eigenstate energy, and $\beta=1/k_BT$ at the temperature of the local environment. The Boltzmann weighted leaking value is therefore defined as
\begin{equation}
L_\text{boltz}=\sum_i w_i \textbf{d}_i^2~.
\end{equation}
Here, the index $i$ is being used to label the oscillator strength in the energy basis rather than the site basis.

\begin{figure}
\centering
	\includegraphics[width=0.8\textwidth]{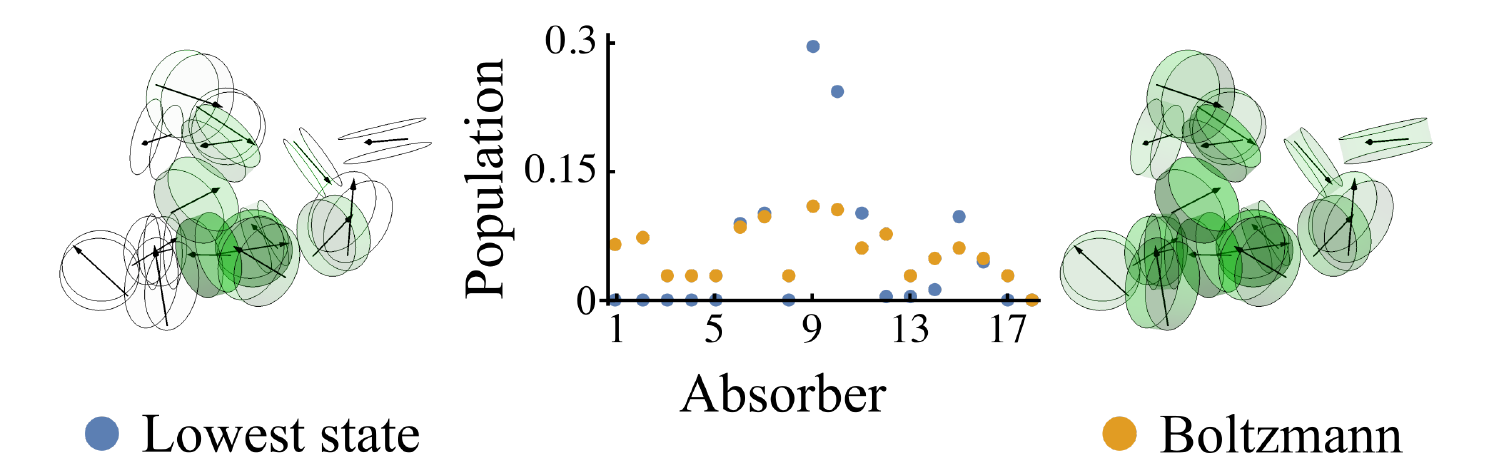}
\caption{The population localisation on the IsiA monomer sites for the steady state found with both the lowest state leaking value and the Boltzmann leaking value at $T=300$~K.}
\label{fig:NoDisLocalisation}
\end{figure}

After having identified both the lowest eigenstate and the Boltzmann-weighted combination of eigenstates that determine the optical brightness of the IsiA monomer, we can transform from the eigenbasis back to the site basis. This allows us to determine the weightings of different chlorophyll A molecules that contribute to the IsiA monomer's optical brightness in its zero temperature or thermalised state, respectively.
In Fig.~\ref{fig:NoDisLocalisation} we show the distribution of population amongst the molecules in a monomer for both the lowest state as well as the Boltzmann leaking value at room temperature. We can see the same molecules take most of the population in both scenarios, but the Boltzmann version flattens out the distribution slightly, as would be expected when more optical pathways are contributing.

\subsection{Introducing disorder}

The nature of our model allows us to introduce disorder in principle to both the optical dipole properties and geometric positions of our molecules, for the latter in a similar manner as has been done in the context of the FMO complex~\cite{Knee2017}. If one were to vary the optical properties of each molecule then the dipole moments of each chromophore would change, in turn affecting the dipole-dipole coupling strengths and similarly leading to a new leaking value. Changing the dipole geometry alters the position and orientation dependent couplings terms between the dipoles, which in turn also affects the eigenstates of the system, and the optical connectivity of those states. 

Whilst the local (protein scaffold) environment will indeed influence the optical properties of the chlorophyll A molecules, in the following we focus on the effects of geometric disorder and keep our dipoles transition energies identical. Following the application of random displacements and rotations to the dipole positions and dipole angles, respectively, we filter the generated geometries to exclude any instances with overlapping disks. This ensures we only keep plausible and consistent disordered configurations. 

We shall refer to our introduced disorder levels in terms of percentages. The percentages represent the standard deviation of the normal distribution used to vary the parameters in each trial. For the position and angle of each 2LS we need to vary 6 parameters, displacements in three Cartesian directions as well as pitch, roll and yaw for the dipole moment. Each parameter needs a characteristic value to take a percentage of to define the standard deviation. Angular disorder uses a $2\pi$~rad complete rotation, while position disorder uses the characteristic length scale defined as the largest distance between two dipoles in the monomer (4.6~nm \cite{Toporik2019}). Due to the three disordered displacements for both position and angle we divide our characteristic values by $\sqrt{3}$. Consequently, whenever we refer to, say, 2\% disorder in the following we shall mean both positional shifts in all three coordinates and three orthogonal rotations all based on Gaussian distributions with, respectively, widths $0.53$~\AA~and angles of $0.023\pi$~rad.

\section{Results}

We start our analysis by calculating the leaking value (representing the brightness of an IsiA monomer benchmarked against a single chlorophyll molecule) based on the original master coordinates of chlorophylls from Ref.~\cite{Toporik2019}. This results in the lowest (single excitation manifold) state leaking value of 0.195, which is a five-fold increase in the ability to hold on to an excitation event. By contrast, the Boltzmann leaking value at 300~K evaluates to 1.29, i.e.~a room temperature lifetime roughly six times shorter than would be expected at temperatures nearing $T=0$ K. This trend is in good qualitative agreement with experimentally obtained results in Ref.~\cite{VanderWeij-deWit2007} which reports a similar factor of $\approx4-5$ separating cryogenic from room temperature fluorescence lifetimes. Note we do not expect full quantitative agreement about absolute values of lifetimes, as we here neither account for non-radiative relaxation nor for carotenoids in the structure. In terms of localisation, the model predicts the highest excitation density occupying chlorophyll 9 in both cases (labeled 509 in the PDB entry and highlighted in Fig.~\ref{fig:chl9}). 

\begin{figure}
\centering
	\includegraphics[width=1\textwidth]{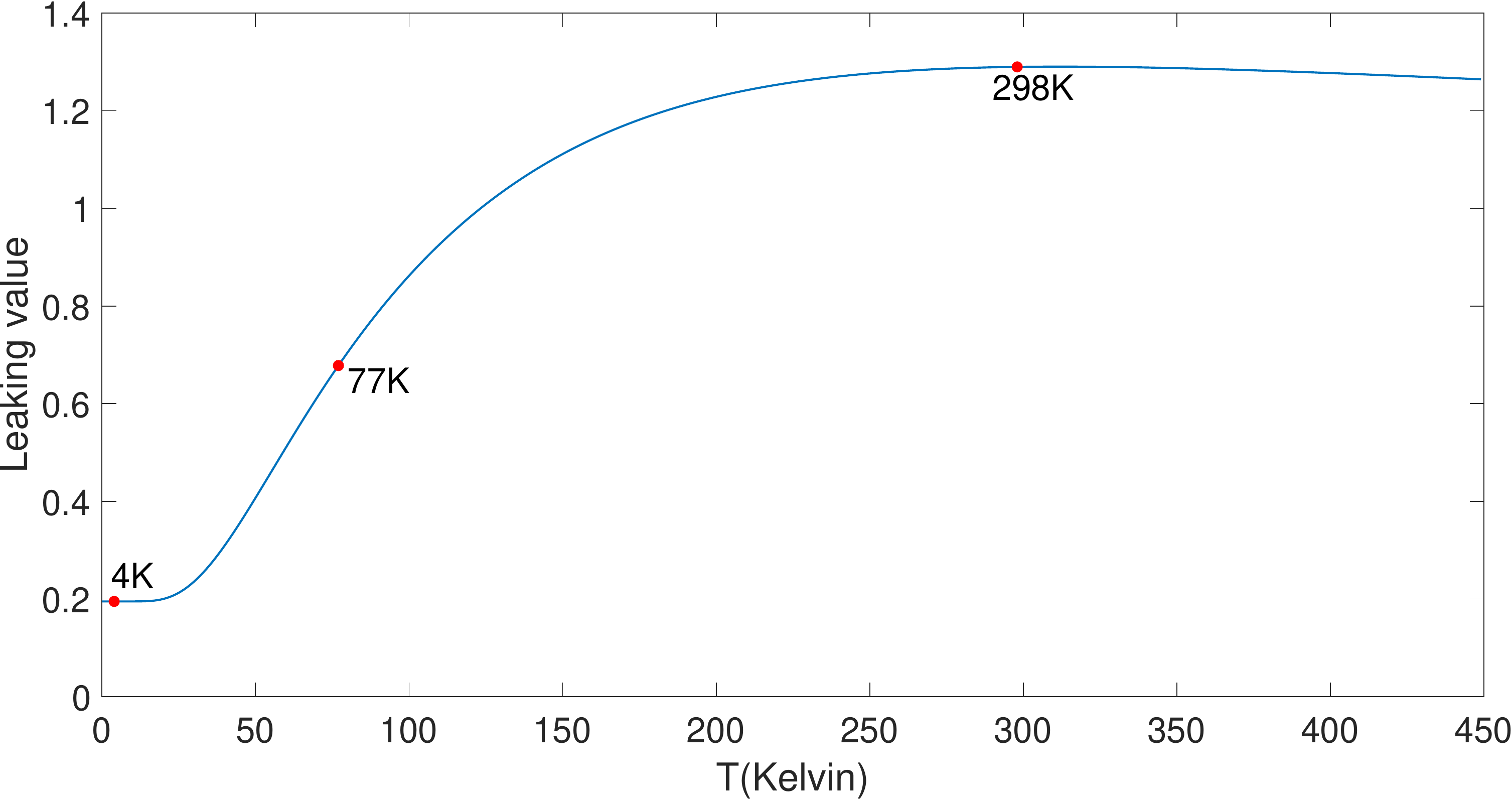}
\caption{Boltzmann leaking value as a function of temperature for an IsiA monomer with all chlorophylls in their master positions. Room temperature, liquid nitrogen and liquid helium highlighted for reference. Note that the limit of $T\to0$~K corresponds to the `lowest state' leaking value.}
\label{fig:IsiABoltzTemp}
\end{figure}

An appreciation of the temperature dependence of the leaking value is important: the lowest-state leaking value is relevant for studies seeking to understand the structure's behavior under cryogenic experimental conditions, whereas the $T=300$~K Boltzmann leaking value applies to the biological structure and function, as biological processes occur, and have evolved, in temperatures of circa 300~K. Figure~\ref{fig:IsiABoltzTemp} shows the leaking value's full dependence on temperature with the experimentally relevant points at liquid helium, liquid nitrogen and room temperature marked for convenience.  We note that the room temperature leaking value occupies a very flat optimum, and no significant change is predicted over a temperature range of $\pm 100$~K, signifying a high degree of robustness. Physically, this temperature insensitivity arises once $k_B T$ dominates over a sufficient number of pairwise energetic spacings between excitonic states, and a higher degree of population equilibration (in the eigenbasis) is approached. 

Encouraged by the fact that our model qualitatively captures the changes in the fluorescence lifetime when compared with measurements in vivo and in vitro~\cite{VanderWeij-deWit2007}, we proceed with an exploration that is not easily achievable experimentally. Namely, we change the given positions of the chlorophylls in the system and gauge the resulting leaking values. Three approaches were taken: randomly shifting all chlorophylls by 1-10\%, randomly shifting each chlorophyll separately by 2\%, and completely removing one chlorophyll at a time. All three approaches gave interesting and insightful results presented below.

\subsection{Removal of selected chlorophylls:}

\begin{figure}
\centering
	\includegraphics[width=\textwidth]{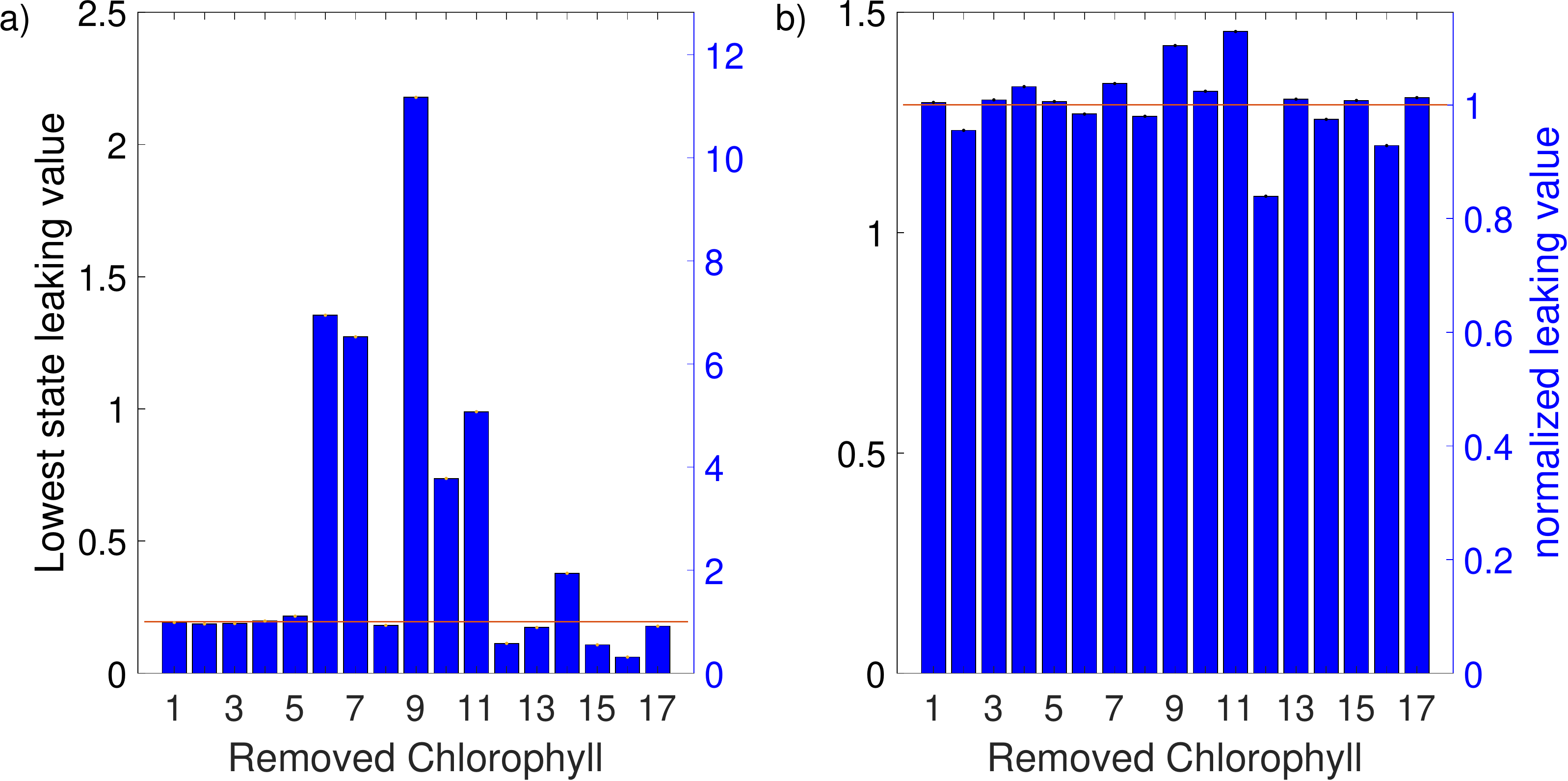}
\caption{Leaking values of structure with one chlorophyll removed. a) Lowest excited state leaking values. b) Boltzmann corrected leaking values. The red line denotes the leaking value of unmodified complete structure. Left (black) axis represents leaking value, right (blue) axis represents leaking value normalised to that of the full structure.}
\label{fig:remove1chlor}
\end{figure}

When considering only the lowest excited state (Fig.~\ref{fig:remove1chlor} a) (relevant at temperatures nearing 0~K) we see some chlorophylls show no or little effect on the leaking value when removed from the master configuration (1,2,3,4,5,8,13, and 17), three that show significant lowering of it (12, 15, and 16) and six which raise it (6,7,9,10,11, and 14 to a lesser extent). When taking into account thermal distribution (Fig.~\ref{fig:remove1chlor} b) major differences are seen in magnitude, and in two cases (chlorophylls 6 and 14) the sign of the change switches from positive to negative.

\subsection{Random shift of single chlorophylls:}

\begin{figure}
\centering
\includegraphics[width=\textwidth]{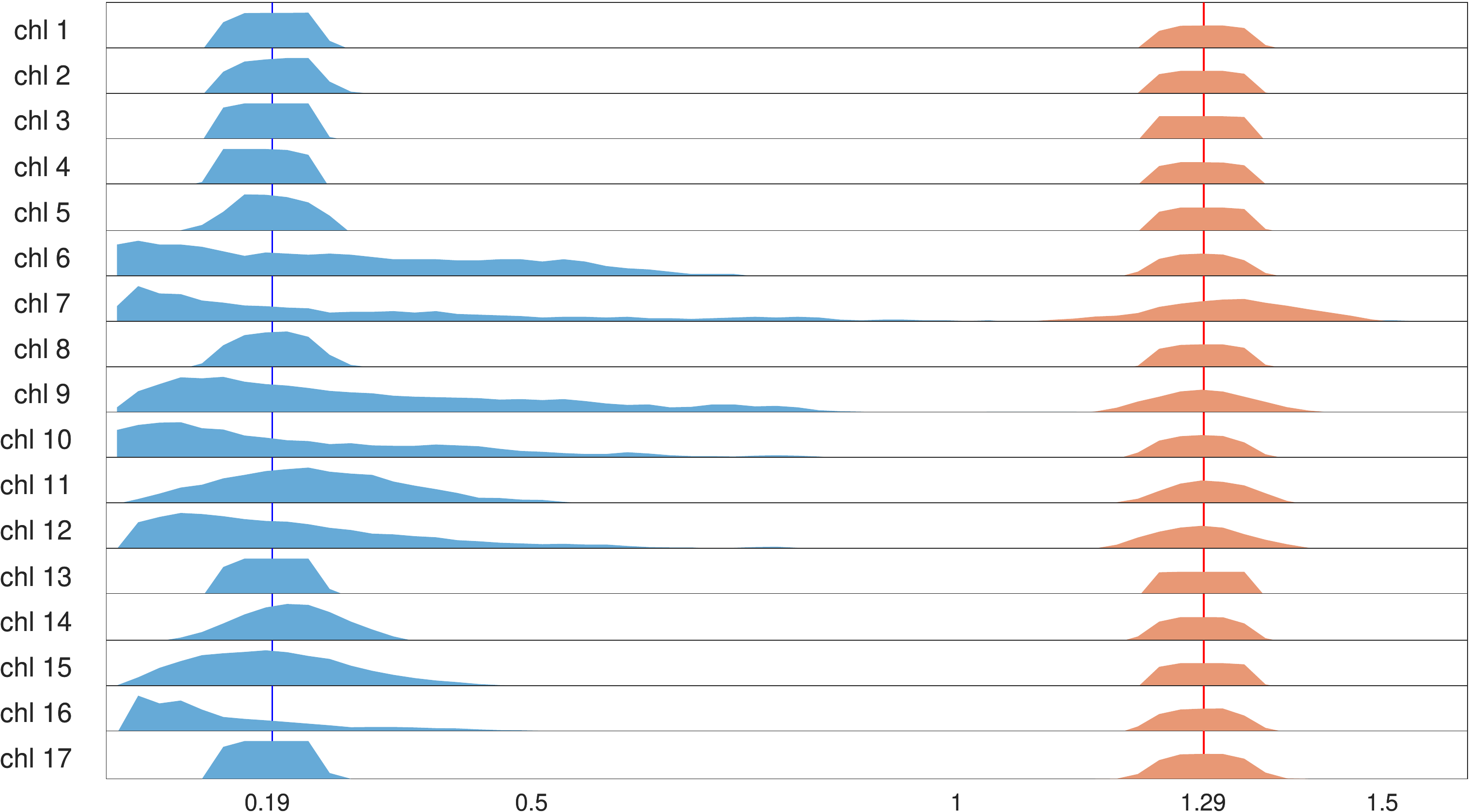}
\caption{Distribution of leaking values. Blue shows the the lowest excitation state leaking value, and red the Boltzmann corrected ones. The vertical lines represent original structure leaking values in both cases. The $y$-axis represents probability, renormalised in each case for best visibility of the respective histogram, and the $x$-axis represents the leaking value. Unity on the $x$-axis is equivalent to a single chlorophyll molecule in solution.}

\label{fig:singleshiftdist}
\end{figure}

\begin{figure}
\centering
	\includegraphics[width=0.8\textwidth]{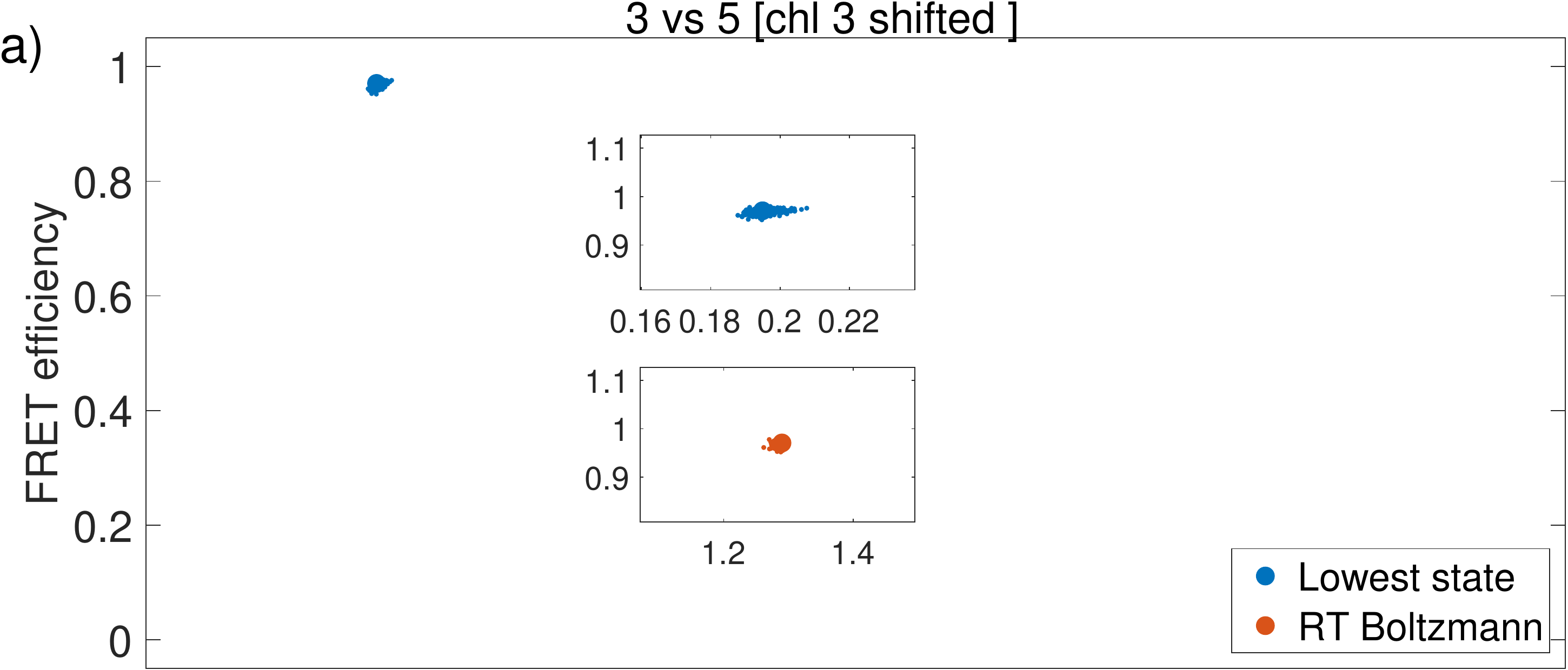}
	\includegraphics[width=0.8\textwidth]{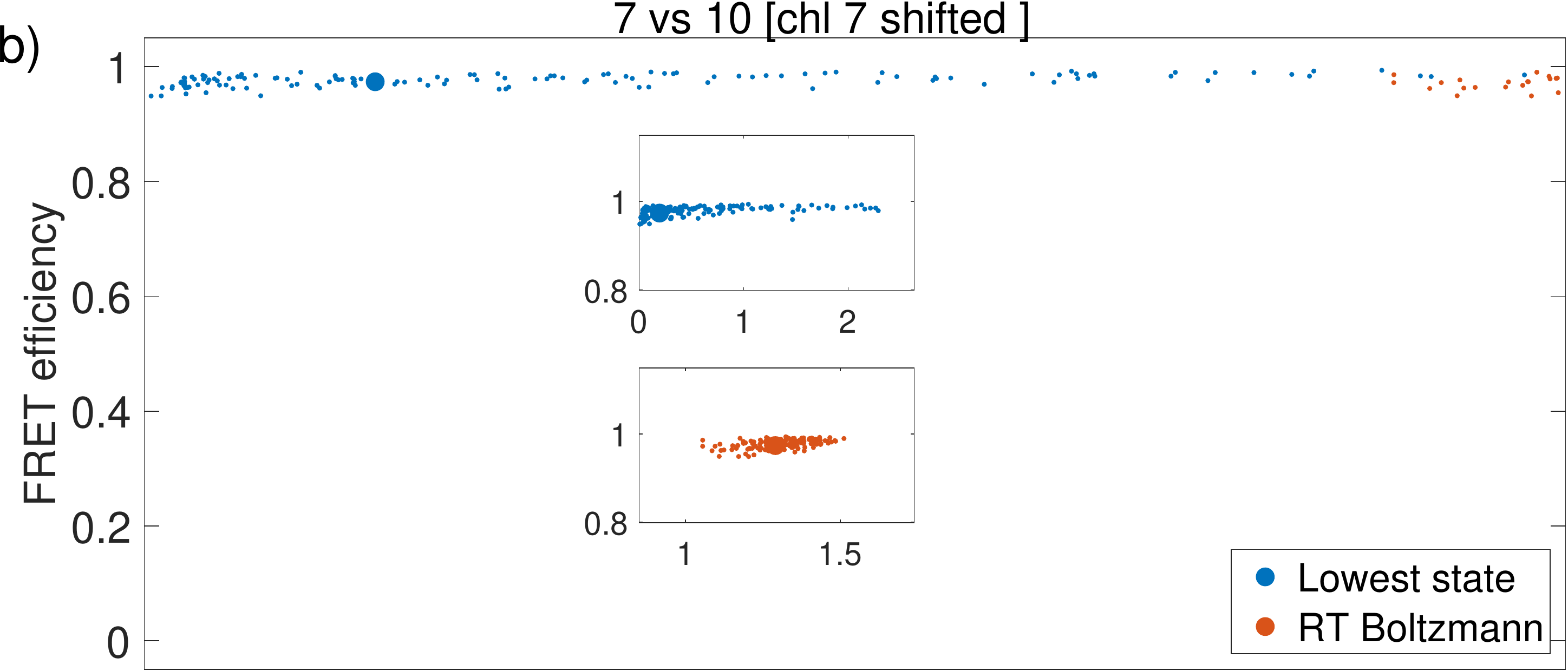}
	\includegraphics[width=0.8\textwidth]{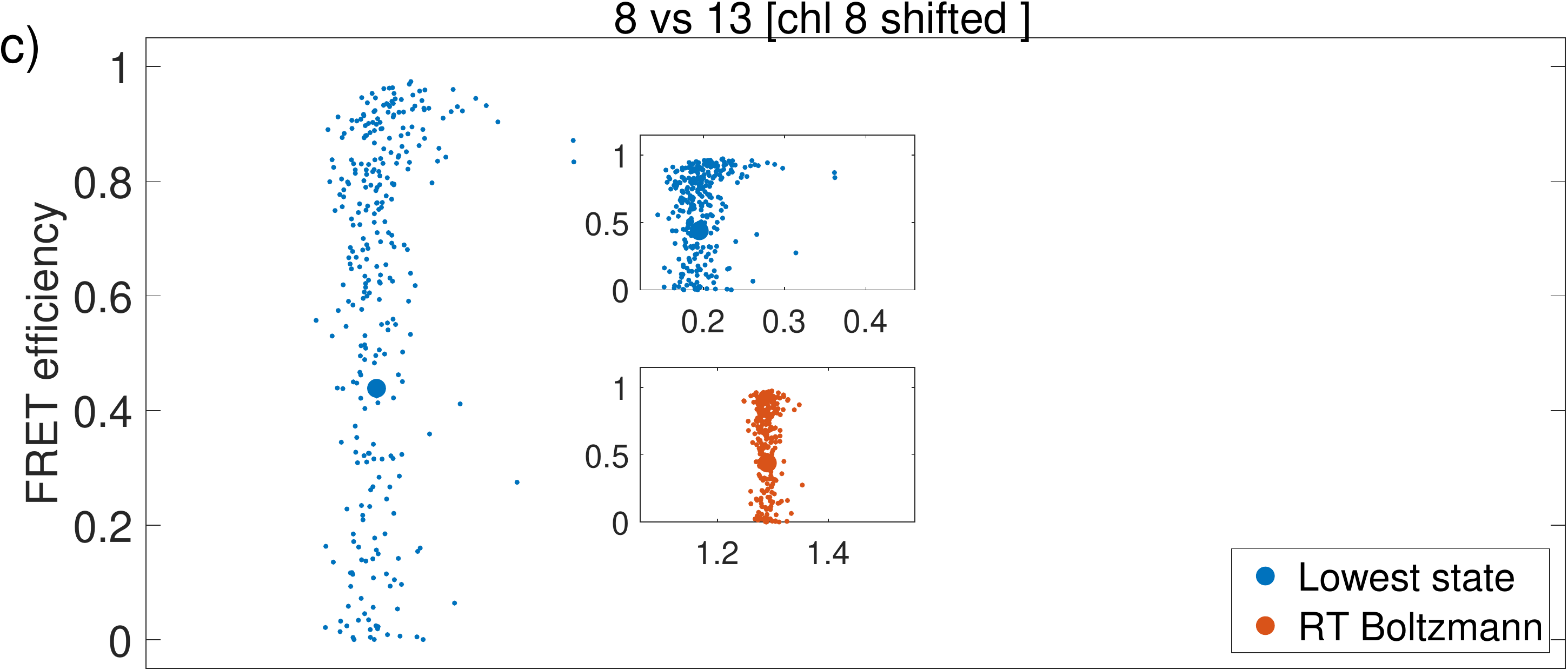}
	\includegraphics[width=0.8\textwidth]{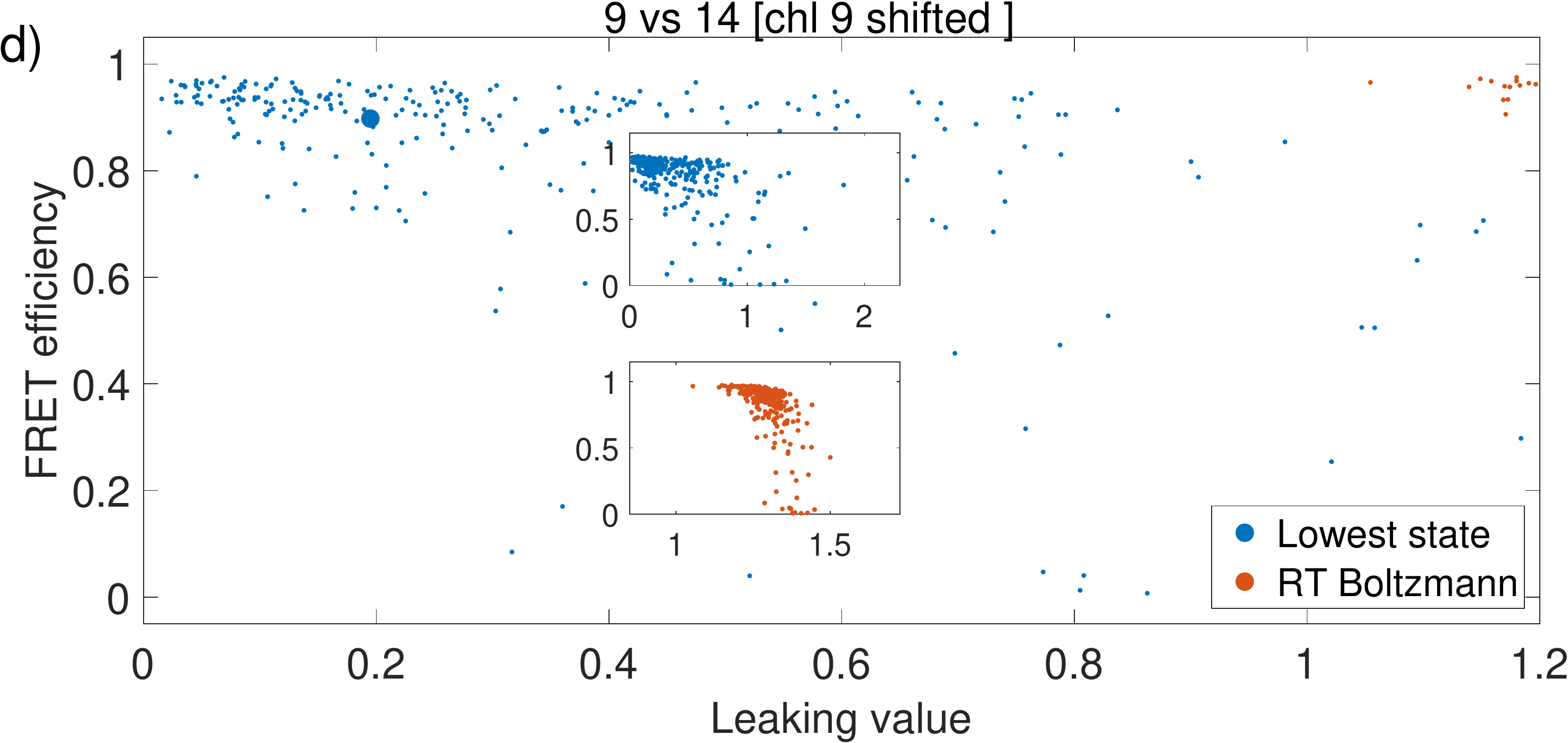}
\caption{Example of types of FRET efficiency vs leaking value patterns we observed (lowest excited state in blue, Boltzmann corrected in red). The zoomed in insets give clearer view of the trends. The four patterns shown are (a) little change to either parameter; (b) very small changes in FRET efficiency shows very large differences in leaking values; (c) large changes in FRET efficiency without corresponding change to leaking value, and (d) both parameters change but no obvious correlation exists between the two.}
\label{fig:fret}
\end{figure}

One at a time, each chlorophyll was shifted in its angular and positional coordinates by a random amount taken from a normal distribution with a 2\% relative disorder. For each chlorophyll between 109 and 299 different positions were considered (with the variability arising from rejection of overlapping geometries). The distributions of the lowest state and Boltzmann leaking values is presented in Fig.~\ref{fig:singleshiftdist}. These results show high variation between the different chlorophylls and between the lowest excited state and Boltzmann corrected results. To try and understand the underlying factors which control the leaking values we plotted the leaking value of each individual run against the FRET efficiency  between each two chlorophylls. Here the FRET efficiencies were taken from Ref.~\cite{Kashida2018}, and this resulted in a total of 272 plots. Analysing these plots we identified four recurring patterns when examining the relationship between the FRET efficiency between pairs of chlorophyll and the overall leaking values, representative examples of which are given in Fig.~\ref{fig:fret}. These four emerging general motifs are:
\begin{itemize}
    \item small variations in FRET efficiency lead to small variations in leaking values;
    \item small variations in FRET efficiency lead to large variations in leaking values;
    \item large variations in FRET efficiency lead to small variations in leaking values;
    \item variations in both FRET efficiency and leaking value are large but uncorrelated.
\end{itemize}
The first three cases are representative of different types of structure to function dynamics: insensitive, highly tunable, and robust. All three are represented in the chlorophyll network within IsiA. While some chlorophylls require very small changes in order to drastically change IsiA's functionality, others can be changed with little consequence. 

\subsection{Random shifting of all chlorophylls:}

\begin{figure}
\centering
	\includegraphics[width=\textwidth]{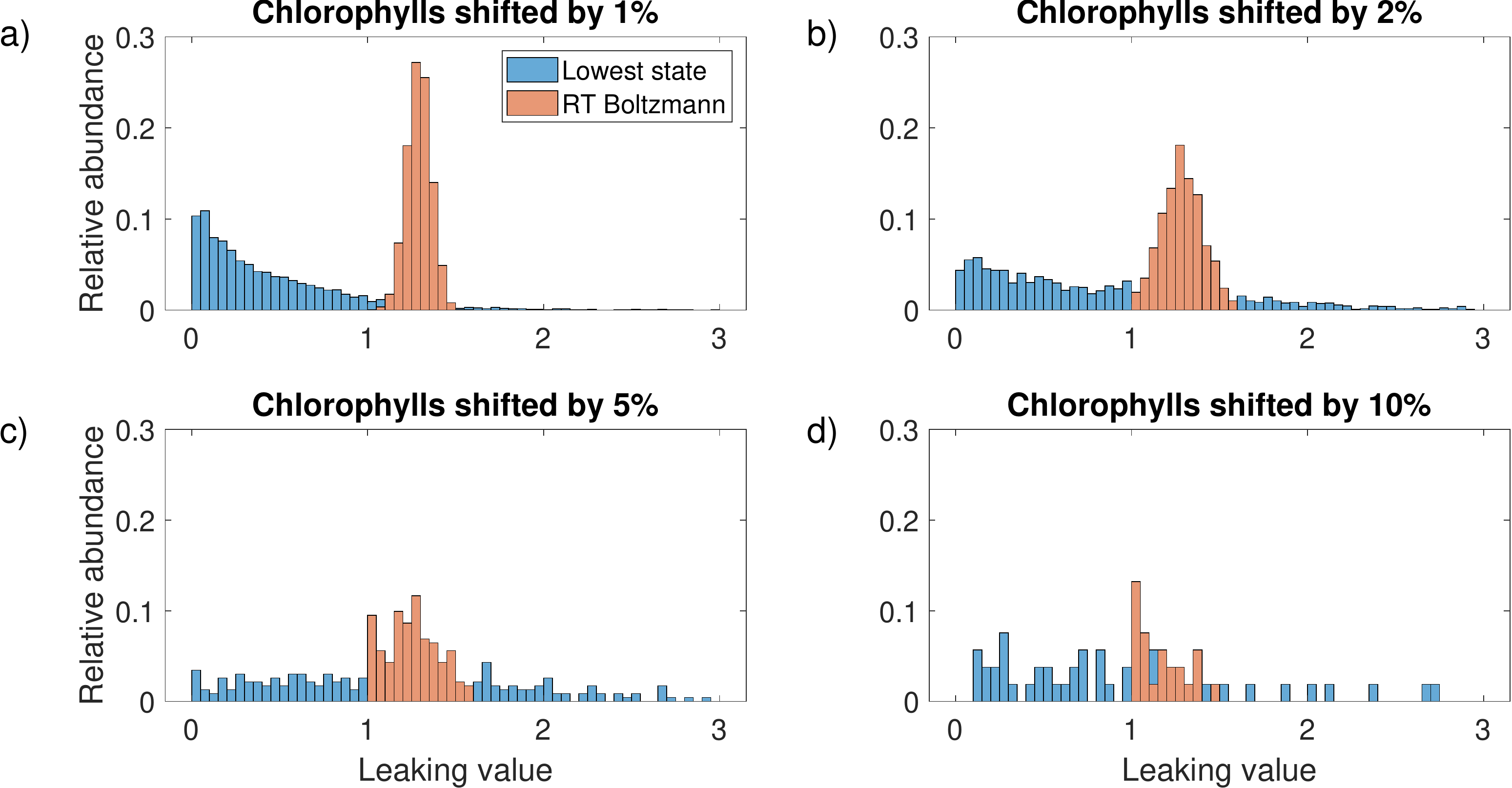}
	\includegraphics[width=\textwidth]{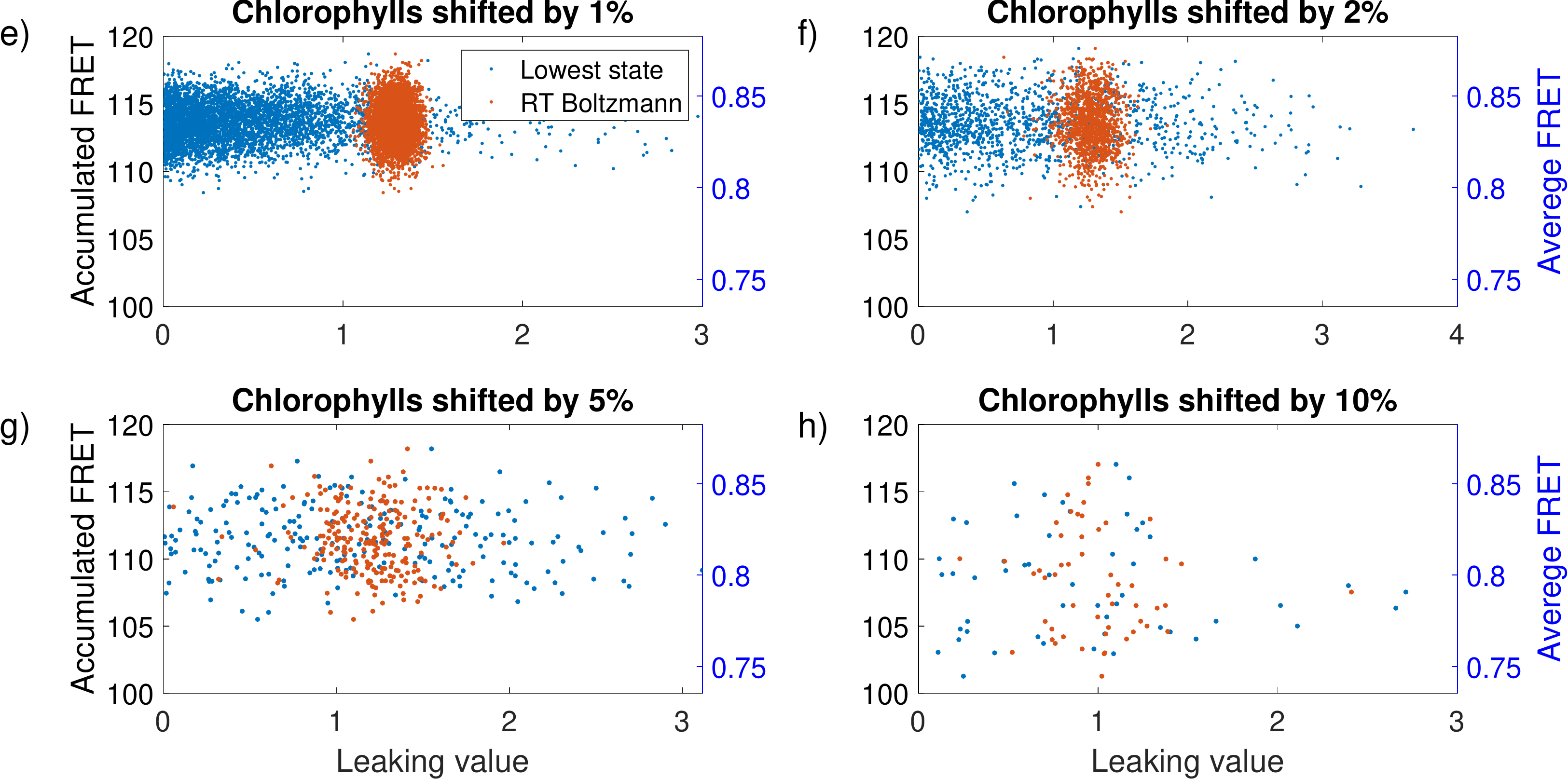}
\caption{Accumulated results of IsiA monomer with introduced variations from the resolved structure. Panels a-d show the distribution of leaking values with introduced 1-10\% change in chlorophyll positions (lowest manifold - blue, room temperature Boltzmann corrected - red). panels e-h show the scatter of leaking values as a function of average FRET efficiency (right axis) and accumulated FRET efficiency (right axis).}
\label{fig:fulldisorder}
\end{figure}

Further runs of the simulation were carried out to gauge IsiA's function and changes thereof when noise is introduced to the system. To examine IsiA's possible dynamic range we introduced normally distributed relative shifts of 1\%, 2\%, 5\% and 10\% (positional and angular) to all chlorophylls in the master configuration simultaneously. The results shown in Fig.~\ref{fig:fulldisorder} show a relative robustness of the structure. Most changes do not lead to very large changes in leaking value, neither at low nor room temperature. Despite the structure's generally robust nature, several specific changes do exist which may entail rather large changes in leaking values in both directions. As long as the disorder level is small, i.e. kept to 1\% - 2\%, the resulting distribution is fairly symmetrical at room temperature (but less so under cryogenic conditions).  

\section{Discussion}

\subsection {Native structure:}

\begin{figure}
	\includegraphics[width=0.4\textwidth]{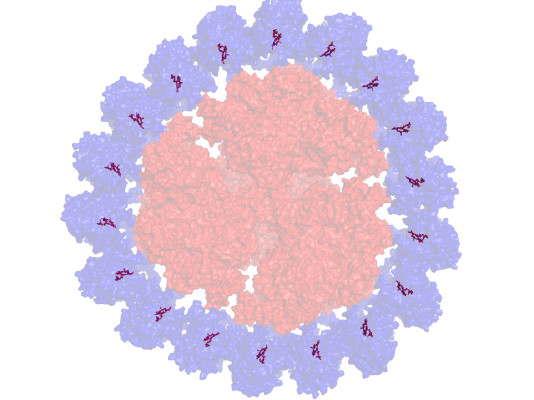}
	\includegraphics[width=0.6\textwidth]{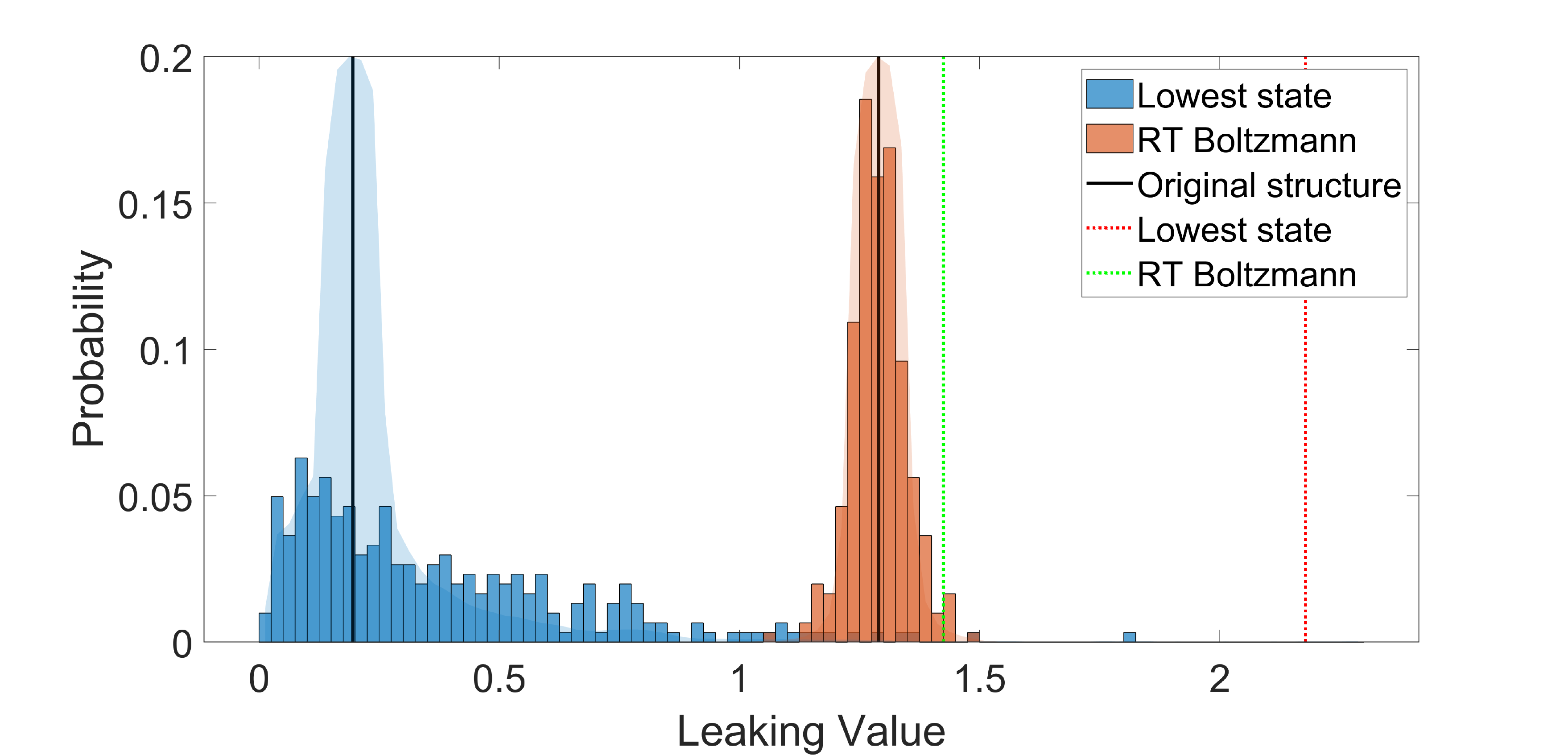}
	\includegraphics[width=0.4\textwidth]{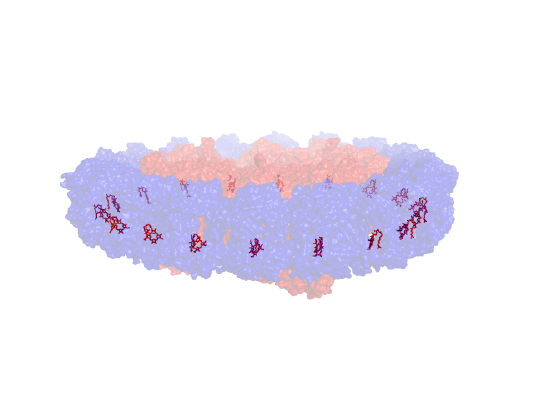}
\caption{Summary of chlorophyll 9 results and images showing the locations of this chlorophyll in the complex: PSI appears in red, IsiA in blue, and chlorophyll 9 in each monomer appears purple. The histogram shows the leaking value of the structure with chlorophyll-9's position randomly shifted and rotated by 2\%. Vertical dashed lines represent the leaking value when chlorophyll 9 is removed. The leaking value of the native structure is represented by black vertical lines. Shaded blue and red curves represent the accumulated results of each of the 17 chlorophylls shifted by 2\% adjusted to highlight the unique contribution of each chlorophylls motion to the overall range of possible leaking values of the IsiA structure. A similar figure was generated for each chlorophyll in the structure (see Supplementary Material).}
\label{fig:chl9}
\end{figure}

Our results for the resolved structure of IsiA here contribute to the current debate about the role of IsiA in connection to its supporting function of the PSI. The central role of, and pronounced localisation of excitations at, chlorophyll 9 does not tally with the notion of IsiA serving as an auxiliary antenna for PSI since chlorophyll 9 is not in the PSI-IsiA interface but rather in the centre of the stromal face (Fig.~\ref{fig:chl9}). This localisation would thus seem more likely to result in higher fluorescence rather than efficient energy transfer. It is worth recalling that while all chlorophylls are chemically identical (and have been modelled as such), their energetic properties may appreciably depend on their protein environment. IsiA associated chlorophylls are known to have a blue-shifted absorption peak in comparison to PSI associated chlorophylls. This shift would seem to indicate the option of a downhill energy transfer form IsiA towards PSI. However, the very strong fluorescence seen in vivo originating from IsiA~\cite{Schoffman2019} fits well with the observation of higher excitation density located farther away from the PSI-IsiA interface.

In vivo, IsiA exhibits a long fluorescence lifetime~\cite{Chen2017}. Neighbouring chlorophylls are located (Mg to Mg) between 1 and 3 nanometres apart. With this proximity and large spectral overlap between emission and absorption the expected F\"{o}rster rates are sub-picosecond. Typically, F\"{o}rster type energy transfer occurs on the picosecond to nanosecond timescale, whereas for many photochemical reactions, the smallest relevant timescale falls into the nanosecond to microsecond range. The ability of the IsiA's chlorophyll network to protect excitations for longer from fluorescent emissions than its individual building blocks may have functional importance. For example, IsiA could function as a buffer mediating between the picosecond timescales of excitation and the photochemical timescales governing charge separation and electron transfer reactions. 

Natural systems operate at temperatures ranging between 270-330~K. At these temperatures, vibrational modes are known to significantly broaden the spectral overlap between pigments, and facilitate fast thermalisation within excitation manifolds, affecting the distribution of exciton population across the molecules. Figure~\ref{fig:IsiABoltzTemp} shows the IsiA monomer's calculated Boltzmann leaking value at different temperatures. Many biophysical measurements are conducted in cryogenic conditions in either liquid nitrogen (77K) or liquid helium (4K). The higher resolution makes these low temperature measurements invaluable for understanding pigment-protein bioenergetics. The large difference both in value and in behaviour between lowest-state and Boltzmann results shown in Figs.~\ref{fig:remove1chlor} through~\ref{fig:fulldisorder} demonstrate the potential for qualitatively vastly differing behaviour between cryogenic temperatures and room temperature. These differences should be considered when assigning biological function based on cryogenic measurements.

\subsection{Removal of selected chlorophylls:}
Removing a co-factor is a classical biological approach for identifying its importance and its role. Theoretically, this grants insights which are easily comparable -- the whole protein with the co-factor vs the protein without the co-factor. However, the experimental process often perturbs more than just the single co-factor it was designed to affect. In addition to that uncertainty, the method of creating a specific mutation is labour-intensive and lengthy. In this study we took a computational approach to probe the possible effects of removing a single chlorophyll. To achieve that we have compared simulations of 17 models each missing one chlorophyll to the full monomer containing all 17 chlorophylls. 
The results in Fig.~\ref{fig:remove1chlor} show a different overall leaking value for more than half of the removed chlorophylls. In some cases the result may be extreme, as in the case of the tenfold increase of leaking value in absence of chlorophyll 9. When comparing the results of the lowest state against the Boltzmann corrected values, the most noticeable difference is the magnitude of change. At room temperature, the removal of any single chlorophyll will result in a 20\% change at most. While still significant, it is much lower than the tenfold increase possible at cryogenic temperatures. Surprisingly, the removal of chlorophyll 6 significantly increased the overall leaking value at cryogenic temperatures while slightly decreasing it at room temperature. These results are interesting from a biological perspective -- during iron limitation the IsiA protein structure is formed and populated with chlorophylls resulting from PSI degradation~\cite{Schoffman2019}. Thermal broadening at room temperature, in this case, would help IsiA maintain relatively stable function even without all 17 of its chlorophylls while avoiding oxidative damage.

\subsection{Random shift of single chlorophylls:}
Most proteins are not rigid structures and the lipid membrane in which IsiA and the photosystems reside is fluid. The movement of the protein structure causes small changes in the protein scaffold, shifting the pigments with it. Larger and more permanent changes occur causing monomers in the ring structure to differ from one another. In the cryo-EM IsiA-PSI supercomplex structure, each IsiA monomer is structurally affected by its interface with the PSI monomer resulting in a threefold symmetry in the IsiA ring i.e.~six slightly different configurations of the IsiA monomers. A third factor is the differences between species -- IsiA can be found in many species of cyanobacteria, differing in primary amino acid sequences. These differences may cause larger shifts than the shifts caused by fluid motion. According to the results shown in Fig.~\ref{fig:singleshiftdist} not all chlorophylls are equally sensitive to such change. While most changes in most chlorophylls do not greatly affect the leaking value, some chlorophylls have a wider distribution and therefore certain specific changes can significantly change the structure's excitonic properties. The changes seen here are reduced at room temperature as compared to the lowest excited state ones. This simulation demonstrates that a pigment protein complex can serve different purposes in a context dependent manner. Such a change can be brought about with only a slight perturbation to a single chlorophyll. While overall IsiA is robust, we can assume that with specific modifications it can shift its function from an antenna to an energy quencher.

\subsection{Random shifting of all chlorophylls:}
Random motion is always present in temperatures relevant to biology and so we can expect bioenergetic processes to function with intrinsic disorder present. We introduced four levels of disorder to our simulation, changing the positions of all chlorophylls at once in a random fashion with shifts of `severity' 1, 2, 5 and 10\% (Fig.~\ref{fig:fulldisorder}). While changes arising from the cases 1\% and 2\% are reasonable, 5\% and 10\% are rather large for random motion within the protein scaffold. Most random shifts within the smaller percentage changes maintain nearly the same leaking value within an even distribution in the lowest excited state. When exerting larger changes, the distribution is wider and is therefore more susceptible to more significant functional variations. When changing all positions simultaneously we cannot attribute changes in the excited lifetime to a specific chlorophyll, however, we can look at the overall average, or accumulated, FRET efficiency within each monomer. While we expect that lower efficiencies correspond to higher leaking values due to the lessened probability of energy transfer, we do not observe this behaviour. This possibly indicates that IsiA is a complex system with redundant excitation pathways. Moreover, we see a narrower distribution when considering the system at room temperature. This difference is especially interesting contrasted against the potential for higher leaking values being greater at cryogenic temperatures. These results line up with both of our previously suggested hypotheses: (\emph{a}) the system has evolved to operate under environmental conditions which include both thermal noise (phonons) and thermally induced disorder (fluid motion) and (\emph{b}) specific changes can result in large effects on functionality. Barring these specific changes, the system is very robust. These observations support the idea of IsiA having context dependent dual function. This data is also consistent with the notion that IsiA, through evolutionary processes, gave rise to pigment protein complexes with different biological functions with a unique chlorophyll configuration which is optimal for the specific function in its specific environment. The notion of the structures' ability to assume different roles is strengthened by the recent works by Cao et al and Akita et al~\cite{Cao_2_2020,Akita2020}: In these studies, the structure of PSI3-IsiA18 from \emph{Synechococcus sp.}~PCC 7942 and from \emph{Thermosynechococcus vulcanus}, respectively, were resolved. Alignment of the two structures revealed subtle differences between the IsiA monomers, as well as differences in the PSI-IsiA interface. This highlights the versatility of the supercomplex structure. These changes may be sufficient to change the energy transfer dynamics both within IsiA and between IsiA and PSI.

\section{Conclusion}
This study supplements experimental data and knowledge concerning the IsiA complex, whose global importance has been extensively discussed in the literature. While experimental studies are invaluable, they contain inherent limitations that do not exist in computational approaches, and this is what we have taken advantage of with this work. This work shows the robust nature of the IsiA complex, the temperature sensitivity of its excitonic landscape, and its potential for change. We believe these aspects add to the current understanding of this complex and we hope they can help direct future experimental endeavours.

Further, we believe the approach taken here, i.e.~combining known biological structures with an efficient numerical modelling, to allow a computational exploration of the relationship between structure and function will be valuable to complement experimental studies in many other contexts.

Finally, as a perspective towards developing artificial light-harvesting devices, our study not only contributes to a better understanding of biological building blocks which can serve as inspiration for such technologies, but also creates and enhances theoretical modelling capability for linking the structure of complex molecular aggregates to energy collection, storage, and transport function.

\section*{Author Contributions}

HS and WMB contributed equally. 
HS and WMB, respectively, developed the computational model of the structure and its physical properties. WMB performed the calculations and HS processed the data. All authors contributed to developing the approach, analysing and interpreting the data, and the writing of the manuscript. NK and EMG supervised the work.

\section*{Acknowledgments}

NK and HS were supported by the ISF grant 1182/19 and the ISF-NSFC grant 2466/18. WMB thanks the EPSRC (grant no. EP/L015110/1) for support. EMG acknowledges support from the Royal Society of Edinburgh and Scottish Government and EPSRC Grant No. EP/T007214/1.

%%%%%%%%%% Insert bibliography here %%%%%%%%%%%%%%

%\printbibliography
%\bibliographystyle{unsrt}
%\bibliography{isiasim2.bib}

\end{document}